\documentclass[aps,prl,twocolumn,showpacs,superscriptaddress,letterpaper,amsmath,amssymb]{revtex4} 

\usepackage{graphicx}  
\usepackage{dcolumn}   
\usepackage{bm}        
\usepackage{amssymb}   
\usepackage{feynmf}
\usepackage{slashed}
\def\gsim{\lower0.5ex\hbox{$\:\buildrel >\over\sim\:$}}
\def\lsim{\lower0.5ex\hbox{$\:\buildrel <\over\sim\:$}}

\newcommand{\be}{\begin{equation}}
\newcommand{\ee}{\end{equation}}
\newcommand{\bea}{\begin{eqnarray}}
\newcommand{\eea}{\end{eqnarray}}

\newcommand{\nbox}{{\,\lower0.9pt\vbox{\hrule \hbox{\vrule height 0.2 cm
\hskip 0.2 cm \vrule height 0.2 cm}\hrule}\,}}

\def\missET {{\not\!\! E_T}}

\begin{document}

\thispagestyle{empty}
\vspace*{-3.5cm}

\vspace{0.5in}


\title{{Search for a heavy vector boson decaying to two gluons \\
in {$\boldmath{p\bar{p}}$} collisions at {$\boldmath{\sqrt{s}=1.96}$ TeV}}}

\affiliation{Institute of Physics, Academia Sinica, Taipei, Taiwan 11529, Republic of China}
\affiliation{Argonne National Laboratory, Argonne, Illinois 60439, USA}
\affiliation{University of Athens, 157 71 Athens, Greece}
\affiliation{Institut de Fisica d'Altes Energies, ICREA, Universitat Autonoma de Barcelona, E-08193, Bellaterra (Barcelona), Spain}
\affiliation{Baylor University, Waco, Texas 76798, USA}
\affiliation{Istituto Nazionale di Fisica Nucleare Bologna, $^{ee}$University of Bologna, I-40127 Bologna, Italy}
\affiliation{University of California, Davis, Davis, California 95616, USA}
\affiliation{University of California, Irvine, Irvine, California 92697, USA}
\affiliation{University of California, Los Angeles, Los Angeles, California 90024, USA}
\affiliation{Instituto de Fisica de Cantabria, CSIC-University of Cantabria, 39005 Santander, Spain}
\affiliation{Carnegie Mellon University, Pittsburgh, Pennsylvania 15213, USA}
\affiliation{Enrico Fermi Institute, University of Chicago, Chicago, Illinois 60637, USA}
\affiliation{Comenius University, 842 48 Bratislava, Slovakia; Institute of Experimental Physics, 040 01 Kosice, Slovakia}
\affiliation{Joint Institute for Nuclear Research, RU-141980 Dubna, Russia}
\affiliation{Duke University, Durham, North Carolina 27708, USA}
\affiliation{Fermi National Accelerator Laboratory, Batavia, Illinois 60510, USA}
\affiliation{University of Florida, Gainesville, Florida 32611, USA}
\affiliation{Laboratori Nazionali di Frascati, Istituto Nazionale di Fisica Nucleare, I-00044 Frascati, Italy}
\affiliation{University of Geneva, CH-1211 Geneva 4, Switzerland}
\affiliation{Glasgow University, Glasgow G12 8QQ, United Kingdom}
\affiliation{Harvard University, Cambridge, Massachusetts 02138, USA}
\affiliation{Division of High Energy Physics, Department of Physics, University of Helsinki and Helsinki Institute of Physics, FIN-00014, Helsinki, Finland}
\affiliation{University of Illinois, Urbana, Illinois 61801, USA}
\affiliation{The Johns Hopkins University, Baltimore, Maryland 21218, USA}
\affiliation{Institut f\"{u}r Experimentelle Kernphysik, Karlsruhe Institute of Technology, D-76131 Karlsruhe, Germany}
\affiliation{Center for High Energy Physics: Kyungpook National University, Daegu 702-701, Korea; Seoul National University, Seoul 151-742, Korea; Sungkyunkwan University, Suwon 440-746, Korea; Korea Institute of Science and Technology Information, Daejeon 305-806, Korea; Chonnam National University, Gwangju 500-757, Korea; Chonbuk National University, Jeonju 561-756, Korea}
\affiliation{Ernest Orlando Lawrence Berkeley National Laboratory, Berkeley, California 94720, USA}
\affiliation{University of Liverpool, Liverpool L69 7ZE, United Kingdom}
\affiliation{University College London, London WC1E 6BT, United Kingdom}
\affiliation{Centro de Investigaciones Energeticas Medioambientales y Tecnologicas, E-28040 Madrid, Spain}
\affiliation{Massachusetts Institute of Technology, Cambridge, Massachusetts 02139, USA}
\affiliation{Institute of Particle Physics: McGill University, Montr\'{e}al, Qu\'{e}bec, Canada H3A~2T8; Simon Fraser University, Burnaby, British Columbia, Canada V5A~1S6; University of Toronto, Toronto, Ontario, Canada M5S~1A7; and TRIUMF, Vancouver, British Columbia, Canada V6T~2A3}
\affiliation{University of Michigan, Ann Arbor, Michigan 48109, USA}
\affiliation{Michigan State University, East Lansing, Michigan 48824, USA}
\affiliation{Institution for Theoretical and Experimental Physics, ITEP, Moscow 117259, Russia}
\affiliation{University of New Mexico, Albuquerque, New Mexico 87131, USA}
\affiliation{The Ohio State University, Columbus, Ohio 43210, USA}
\affiliation{Okayama University, Okayama 700-8530, Japan}
\affiliation{Osaka City University, Osaka 588, Japan}
\affiliation{University of Oxford, Oxford OX1 3RH, United Kingdom}
\affiliation{Istituto Nazionale di Fisica Nucleare, Sezione di Padova-Trento, $^{ff}$University of Padova, I-35131 Padova, Italy}
\affiliation{University of Pennsylvania, Philadelphia, Pennsylvania 19104, USA}
\affiliation{Istituto Nazionale di Fisica Nucleare Pisa, $^{gg}$University of Pisa, $^{hh}$University of Siena and $^{ii}$Scuola Normale Superiore, I-56127 Pisa, Italy}
\affiliation{University of Pittsburgh, Pittsburgh, Pennsylvania 15260, USA}
\affiliation{Purdue University, West Lafayette, Indiana 47907, USA}
\affiliation{University of Rochester, Rochester, New York 14627, USA}
\affiliation{The Rockefeller University, New York, New York 10065, USA}
\affiliation{Istituto Nazionale di Fisica Nucleare, Sezione di Roma 1, $^{jj}$Sapienza Universit\`{a} di Roma, I-00185 Roma, Italy}
\affiliation{Rutgers University, Piscataway, New Jersey 08855, USA}
\affiliation{Texas A\&M University, College Station, Texas 77843, USA}
\affiliation{Istituto Nazionale di Fisica Nucleare Trieste/Udine, I-34100 Trieste, $^{kk}$University of Udine, I-33100 Udine, Italy}
\affiliation{University of Tsukuba, Tsukuba, Ibaraki 305, Japan}
\affiliation{Tufts University, Medford, Massachusetts 02155, USA}
\affiliation{University of Virginia, Charlottesville, Virginia 22906, USA}
\affiliation{Waseda University, Tokyo 169, Japan}
\affiliation{Wayne State University, Detroit, Michigan 48201, USA}
\affiliation{University of Wisconsin, Madison, Wisconsin 53706, USA}
\affiliation{Yale University, New Haven, Connecticut 06520, USA}

\author{T.~Aaltonen}
\affiliation{Division of High Energy Physics, Department of Physics,
University of Helsinki and Helsinki Institute of Physics, FIN-00014,
Helsinki, Finland}
\author{J.~Adelman}
\affiliation{Yale University, New Haven, Connecticut 06520, USA}
\author{B.~\'{A}lvarez~Gonz\'{a}lez$^z$}
\affiliation{Instituto de Fisica de Cantabria, CSIC-University of Cantabria, 39005 Santander, Spain}
\author{S.~Amerio}
\affiliation{Istituto Nazionale di Fisica Nucleare, Sezione di Padova-Trento, $^{ff}$University of Padova, I-35131 Padova, Italy}
\author{D.~Amidei}
\affiliation{University of Michigan, Ann Arbor, Michigan 48109, USA}
\author{A.~Anastassov$^x$}
\affiliation{Fermi National Accelerator Laboratory, Batavia, Illinois 60510, USA}
\author{A.~Annovi}
\affiliation{Laboratori Nazionali di Frascati, Istituto Nazionale di Fisica Nucleare, I-00044 Frascati, Italy}
\author{J.~Antos}
\affiliation{Comenius University, 842 48 Bratislava, Slovakia; Institute of Experimental Physics, 040 01 Kosice, Slovakia}
\author{G.~Apollinari}
\affiliation{Fermi National Accelerator Laboratory, Batavia, Illinois 60510, USA}
\author{J.A.~Appel}
\affiliation{Fermi National Accelerator Laboratory, Batavia, Illinois 60510, USA}
% ...

\collaboration{CDF Collaboration\footnote{With visitors from
$^a$Istituto Nazionale di Fisica Nucleare, Sezione di Cagliari, 09042 Monserrato (Cagliari), Italy,
$^b$University of CA Irvine, Irvine, CA 92697, USA,
$^c$University of CA Santa Barbara, Santa Barbara, CA 93106, USA,
$^d$University of CA Santa Cruz, Santa Cruz, CA 95064, USA,
$^e$Institute of Physics, Academy of Sciences of the Czech Republic, Czech Republic,
$^f$CERN, CH-1211 Geneva, Switzerland,
$^g$Cornell University, Ithaca, NY 14853, USA,
$^h$University of Cyprus, Nicosia CY-1678, Cyprus,
$^i$Office of Science, U.S. Department of Energy, Washington, DC 20585, USA,
$^j$University College Dublin, Dublin 4, Ireland,
$^k$ETH, 8092 Zurich, Switzerland,
$^l$University of Fukui, Fukui City, Fukui Prefecture, Japan 910-0017,
$^m$Universidad Iberoamericana, Mexico D.F., Mexico,
$^n$University of Iowa, Iowa City, IA 52242, USA,
$^o$Kinki University, Higashi-Osaka City, Japan 577-8502,
$^p$Kansas State University, Manhattan, KS 66506, USA,
$^q$Ewha Womans University, Seoul, 120-750, Korea,
$^r$University of Manchester, Manchester M13 9PL, United Kingdom,
$^s$Queen Mary, University of London, London, E1 4NS, United Kingdom,
$^t$University of Melbourne, Victoria 3010, Australia,
$^u$Muons, Inc., Batavia, IL 60510, USA,
$^v$Nagasaki Institute of Applied Science, Nagasaki, Japan,
$^w$National Research Nuclear University, Moscow, Russia,
$^x$Northwestern University, Evanston, IL 60208, USA,
$^y$University of Notre Dame, Notre Dame, IN 46556, USA,
$^z$Universidad de Oviedo, E-33007 Oviedo, Spain,
$^{aa}$CNRS-IN2P3, Paris, F-75205 France,
$^{bb}$Texas Tech University, Lubbock, TX 79609, USA,
$^{cc}$Universidad Tecnica Federico Santa Maria, 110v Valparaiso, Chile,
$^{dd}$Yarmouk University, Irbid 211-63, Jordan,
}}
\noaffiliation

\begin{center}
\begin{abstract}
We present a search for a new heavy vector boson $Z'$ that decays to
 gluons. Decays to on-shell gluons are suppressed, leading to a
 dominant decay mode of $Z' \rightarrow g^*g$. We study the case where the off-shell gluon $g^*$ converts to a pair of top quarks,
 leading to a final state of $t\bar{t}g$. In a
 sample of events with exactly one charged lepton, large missing
 transverse momentum and at least five jets, corresponding to an integrated
 luminosity of 8.7 fb$^{-1}$ collected by the CDF II detector, we find the data to be consistent with the 
 standard model. We set upper limits on the production cross section 
 times branching ratio
 of this chromophilic $Z'$ at 95\% confidence level 
  from 300 fb to 40 fb
 for $Z'$ masses ranging from $400$ GeV$/c^2$ to $1000$ GeV$/c^2$, respectively.
\end{abstract}
\end{center}

\maketitle

\begin{figure}[floatfix]
\includegraphics[width=2.5in]{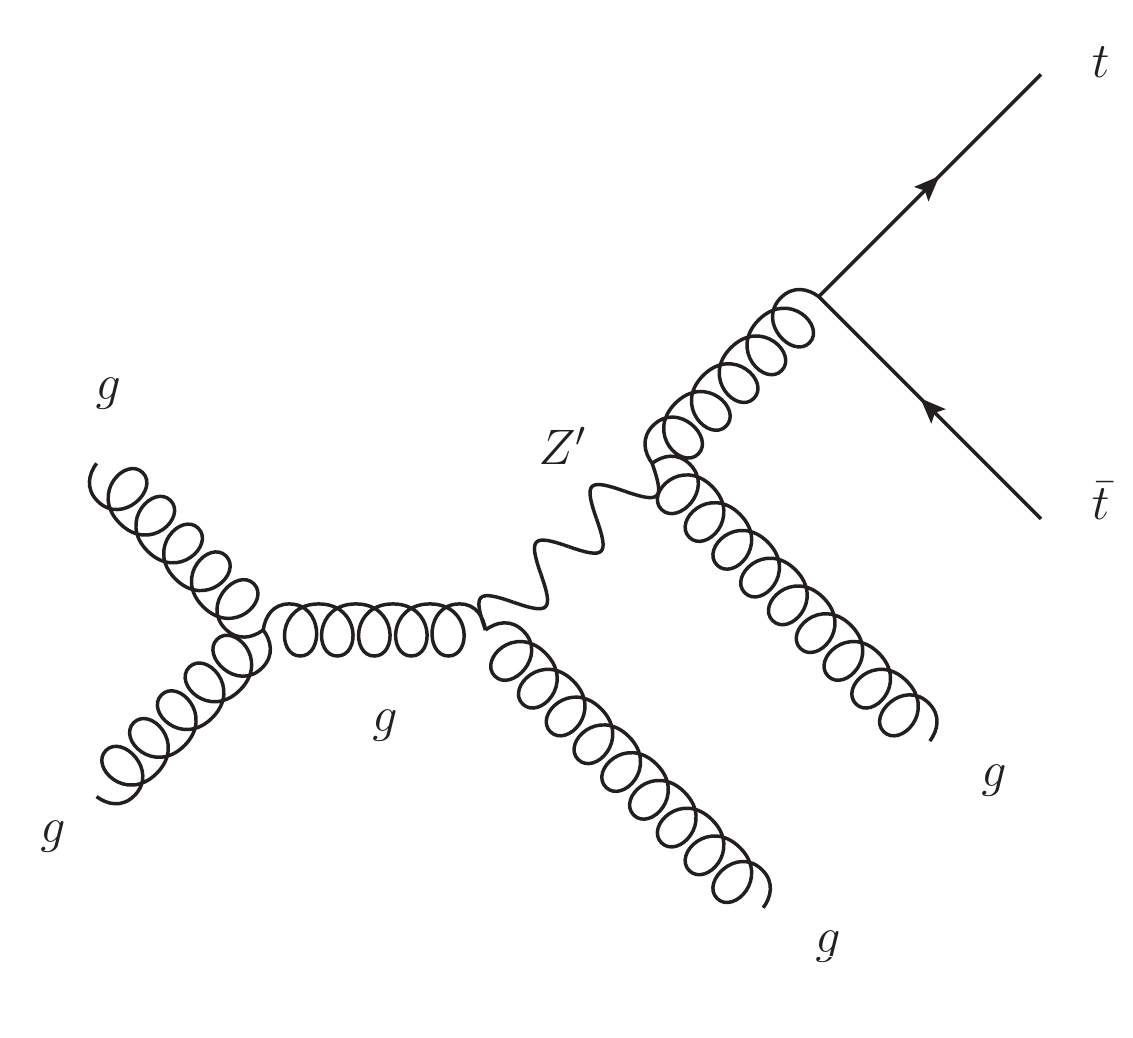}
\includegraphics[width=2.5in]{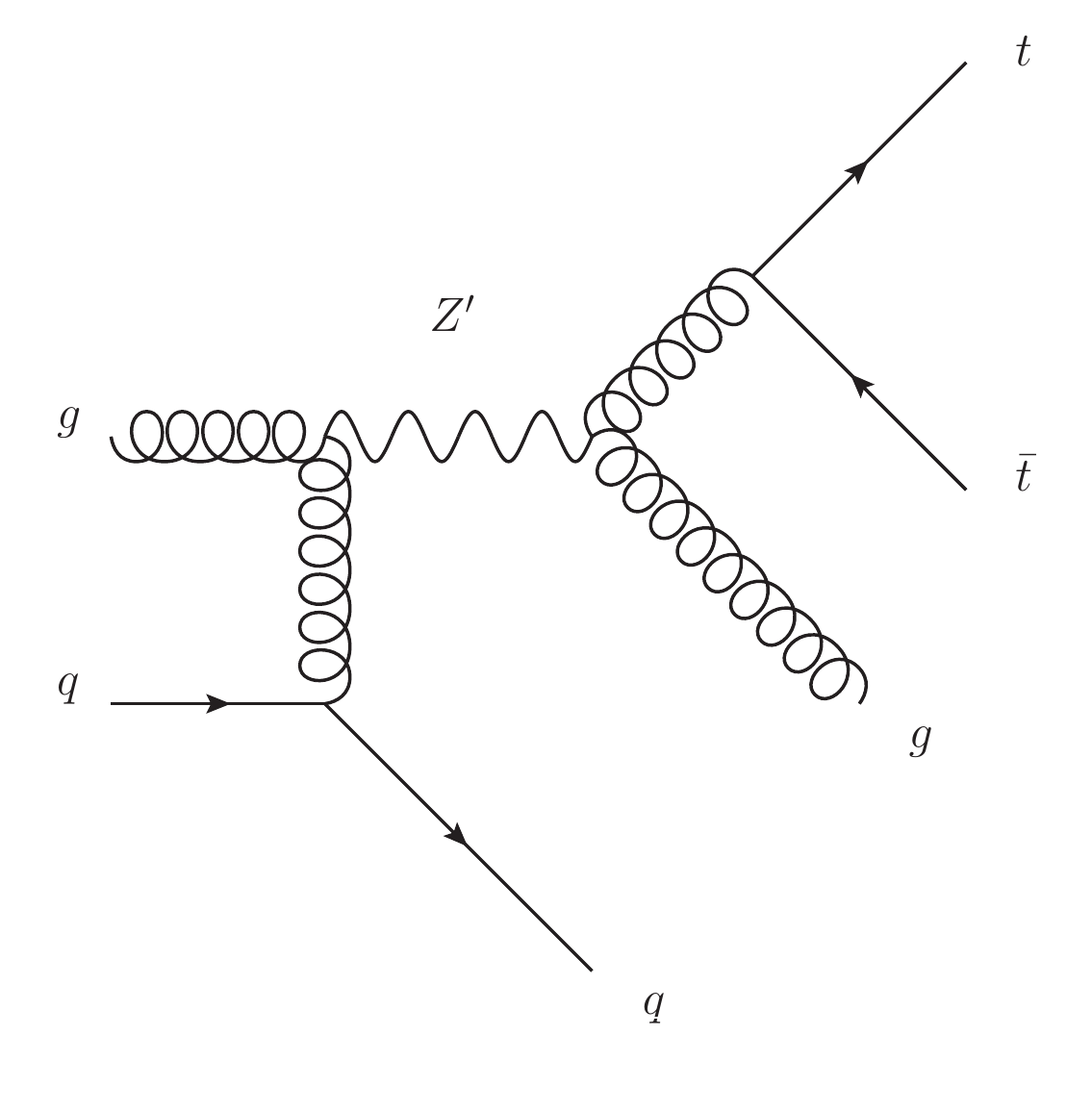}
\caption{ Diagram for $Z'g$ (top) or $Z'q$ (bottom) production followed by $Z'\rightarrow gg^*
  \rightarrow g t\bar{t}$ decay giving a $t\bar{t}gg$ (top) or
  $t\bar{t}gq$ (bottom) final state.}
\label{fig:diagram}
\end{figure}

Various models of physics beyond the standard model (SM) predict new U(1) symmetries with an associated electrically neutral $Z'$ gauge boson. 
Assuming coupling to charged lepton pairs, experiments at the LHC rule out such particles up 
to masses of several TeV~\cite{atlaszp,cmszp}.
Strict limits are also set by D0, CDF, ATLAS and CMS in searches for $Z'$ decaying  to light quarks~\cite{d0zlightq,cdfzlightq,atlaszlightq,cmszlightq} or $t\bar{t}$ pairs~\cite{cdfzpttbar,atlaszpttbar,cmszpttbar,d0zpttbar}. If the 
new particle decays only to gluons (\emph{chromophilic} $Z'$), such limits are 
evaded.  If the new gauge boson is due to a new hidden sector,
tree-level couplings to fermions may be suppressed, and the leading interactions would be with fields charged under the new U(1) and SU(2)
or SU(3) groups; the SU(3) case leads to a  chromophilic
$Z'$ that decays to pairs of gluons~\cite{ourzp}.  However, the
Landau-Yang theorem~\cite{LandauYang} prevents a vector particle from
decaying to two massless gauge bosons, and so the predominant decay
mode is to one on-shell (massless) gluon and one off-shell (massive)
gluon, the latter then decaying to a pair of quarks, giving $Z'
\rightarrow g^*g \rightarrow q\bar{q}g$.  For the same reason, the
$Z'$ boson can not be produced through the fusion of two on-shell
gluons in the process $gg\to Z'$, but require at least one of the
incoming gluons to be offshell, see Fig.~\ref{fig:diagram}.

 If the $g^*\rightarrow q\bar{q}$ pair in the decay is below the
 top-quark pair mass threshold, it gives a four-jet final
state; the usual constraints on $Z'$ models from dilepton and dijet final states
therefore do not apply to this model.  However, the four-jet final
state with a resonance in three jets would be  challenging to see over
the large multi-jet background.  To extract the signal from the large
background, we will look at signal events where the off-shell gluon
decays to heavy flavor quarks. In this paper, we focus on
the decay  $Z'\rightarrow gt\bar{t}$ and  consider the 
decay mode $Z' \rightarrow t\bar{t}g \rightarrow
W^{+}bW^{-}\bar{b}g$ in which one $W$ boson
decays leptonically (including  $\tau$ lepton decays) and the
second $W$ boson
decays  to a quark-antiquark pair. This decay mode features a large $t\bar{t}$
branching ratio and a distinctive experimental signature which allows
the reduction  to a manageable level of the backgrounds
other than SM $t\bar{t}$ production. Such a signal is similar to SM top-quark
pair production and decay, but with an additional jet coming from the
on-shell gluon from $Z'\rightarrow g^*g$ decay.

We analyze a sample of events corresponding to an integrated
luminosity of 8.7$\pm 0.5$ fb$^{-1}$ recorded by the CDF II
detector~\cite{cdf}, a general purpose detector designed to study
$p\overline{p}$ collisions at $\sqrt{s}=1.96$ TeV produced by the Fermilab Tevatron collider. 
CDF's tracking system consists of a silicon microstrip tracker and a
drift chamber that are immersed in a 1.4 T axial magnetic field~\cite{trackingsystems}. 
Electromagnetic and hadronic calorimeters surrounding
the tracking system measure particle energies, with muon detection
provided by an additional system of drift chambers
located outside the calorimeters.

Events are selected online (triggered) by the requirement of an $e$ or $\mu$ candidate~\cite{leptonid} with
transverse momentum $p_T$~\cite{coordinates} greater than $18$ GeV$/c$.
After trigger selection, events are retained if the electron or muon candidate has a pseudorapidity $|\eta| < 1.1$~\cite{coordinates}, $p_T > 20$ GeV$/c$ and satisfies the standard CDF identification and
isolation requirements~\cite{leptonid}. 
We reconstruct jets in the
calorimeter using the {\sc jetclu}~\cite{jetclu} algorithm with a
clustering radius of 0.4 in $\eta-\phi$ space, and calibrated using the techniques outlined in Ref.~\cite{jes}. Jets are required to have transverse energy $E_T > 15$ GeV and $|\eta| < 2.4$. Missing transverse momentum~\cite{metdef}
is reconstructed using 
calorimeter and muon information~\cite{leptonid}; in this experimental
signature
the missing transverse momentum is mostly due to the neutrino from the
leptonically-decaying $W$ boson.

The signature of $Z' \rightarrow t\bar{t}g \rightarrow W^{+}bW^{-}\bar{b}g
\rightarrow  \ell\nu bqq'\bar{b}g$  is a charged lepton ($e$ or
$\mu$), large missing
transverse momentum, two jets arising from $b$ quarks, and three
additional jets from the $W$-boson hadronic decay and the $Z'$ decay gluon.  We select events with exactly one electron or
muon, at least five jets, and missing transverse momentum greater than
20~GeV$/c$. Since such a signal would have two jets originating from
$b$ quarks,
we require (with minimal loss of efficiency) evidence of decay of a
$b$ hadron in at least one jet. This requirement, called $b$-tagging,
makes use of the {\sc secvtx} algorithm, which identifies jets from
$b$ quarks via their secondary vertices~\cite{secvtx}.

We model the production of  $Z'$ bosons with $m_{Z'} = 400$--$1000$
GeV$/c^2$ in 100 GeV$/c^2$ intervals and subsequent decays $Z'\rightarrow gg^*$ and
$g^*\rightarrow t\bar{t}$  with {\sc Madgraph}~\cite{madgraph}. Additional radiation, hadronization and
showering are described by {\sc pythia}~\cite{pythia}. The detector
response for all simulated samples is modeled by the {\sc geant}-based CDF II detector simulation~\cite{cdfsimul}.

The dominant SM background to the $t\bar{t}+j$ signature is top-quark
pair production with an additional jet due to initial-state or
final-state radiation. We model this background using
{\sc pythia} $t\bar{t}$ production with a top-quark mass $m_t = 172.5$ GeV/$c^2$~\cite{topmass}. We
normalize the $t\bar{t}$ background to the theoretical calculation at next-to-next-to-leading order (NNLO) in $\alpha_s$~\cite{nloa}.  In addition,
events generated by a next-to-leading order generator, {\sc mc@nlo}~\cite{mcatnlo}, are used in estimating an uncertainty
	    in modeling the radiation of an additional jet.

 The second-largest SM background process is the associated production 
of a $W$ boson and jets. Samples of  $W$-boson+jets events with
light-flavor and heavy-flavor ($b,c$) quark
jets are generated using {\sc alpgen}~\cite{alpgen},
and interfaced with a parton-shower model 
using {\sc pythia}.  The $W$-boson+jets samples are normalized to the
measured $W$-boson production cross section, 
with an additional multiplicative factor for the relative contribution of heavy- 
and light-flavor jets, following the same technique utilized previously in measuring the
top-quark pair-production cross section~\cite{secvtx}.

To check the quality of the
 $W$-boson+jets background modeling, 
we compare the model to the data in $W$-boson+four-jet events with
zero $b$-tags. These events are expected to contain only 1\% of signal, while $W$-boson+jets events are expected to account for 50\% of the expected background 
yield.  We find agreement between our total background estimate and
the data to within 1\%. 

 Backgrounds due
to production of a $Z$ boson with additional jets, where the second
lepton from the $Z$-boson decay is not reconstructed, are small
compared to the $W$-boson background and are modeled using events
generated with {\sc alpgen}, and interfaced with the parton-shower model 
using {\sc pythia}. The multi-jet
background, in which a jet is
misreconstructed as a lepton, is modeled using events triggered on
jets below the selection threshold
 normalized
to a background-dominated region at low missing transverse momentum
where the multi-jet background is large.

The SM background due to single-top-quark is modeled
using {\sc Madgraph} interfaced with {\sc pythia} parton-shower
models; backgrounds from diboson production are modeled using {\sc
  pythia}. Both are normalized to next-to-leading-order
cross sections~\cite{ewknlo2,ewknlo}. 

\begin{figure}[htp]
\includegraphics[width=0.7\linewidth]{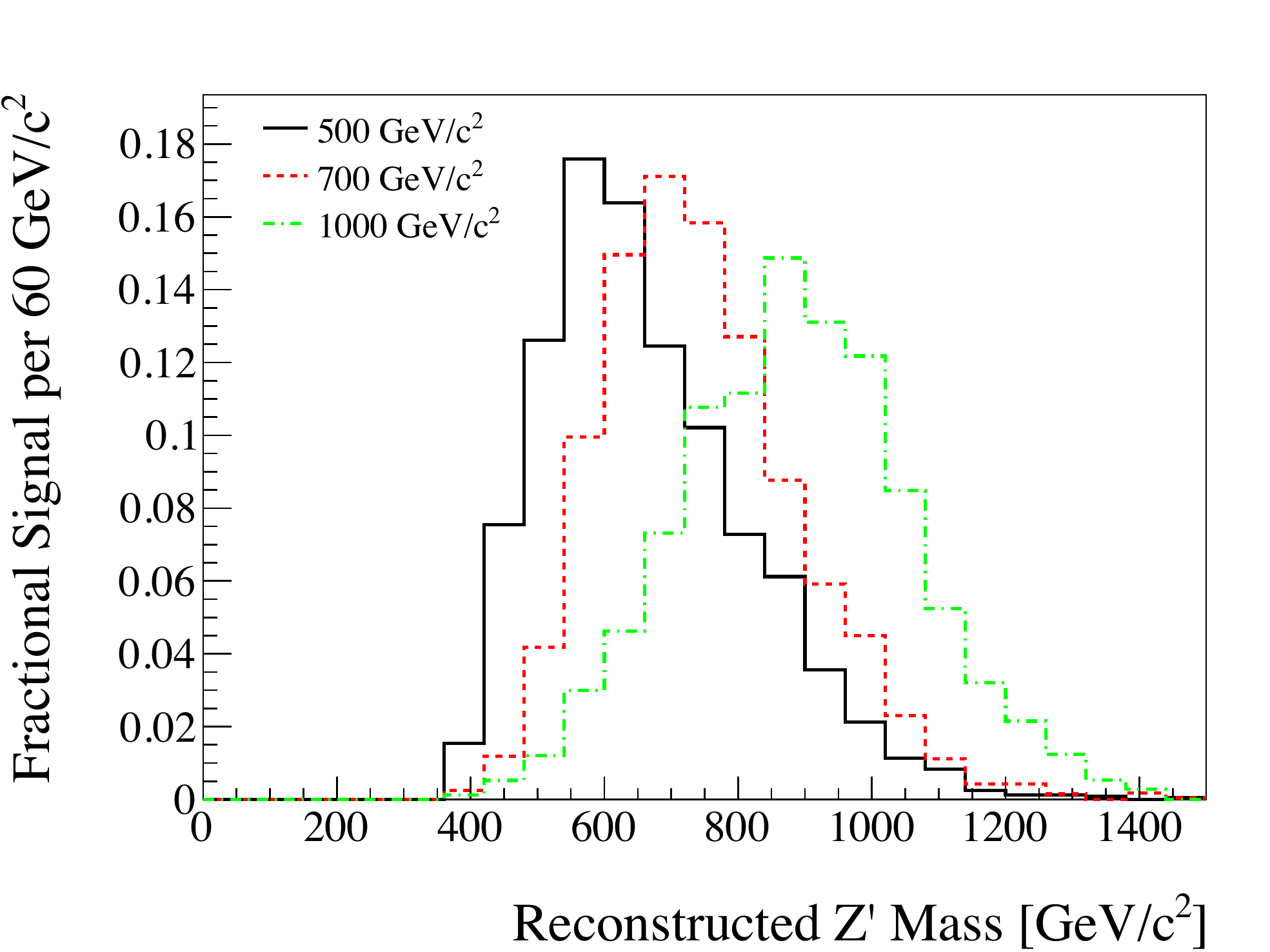}\\
\caption{Distribution of reconstructed $Z'$ mass in simulated events
  for three choices of $m_{Z'}$.}
\label{fig:reco}
\end{figure}

We search for a signal as an excess of events above expectations
from backgrounds in event distributions versus the mass of the $t\bar{t}j$
system ($Z'\rightarrow t\bar{t}j$). In \mbox{$t\bar{t}+j$} events,
 we first identify the jets belonging to the $t\bar{t}$ system, using a kinematic fitter~\cite{rad}
to select from all available jets in the event the four jets most consistent with the $t\bar{t}$ topology.  In the fit, the top-quark and $W$-boson
masses are constrained to be $172.5$ GeV/$c^2$ and $80.4$ GeV/$c^2$,
respectively. All remaining
jets are considered candidates for the light-quark jet in the $t\bar{t}j$ 
resonance.  Following the strategy proposed in Ref.~\cite{ourzp}, we
choose the jet with the largest value of $\Delta R(j,t\bar{t}) \times
p_{\rm T}^{\rm jet}$ to reconstruct the resonance mass
$m_{t\bar{t}j}$, where $\Delta R(j,t\bar{t})$ is the distance between
a jet and the $t\bar{t}$ system in $\eta-\phi$ space.
Figure~\ref{fig:reco} shows distributions of the reconstructed mass for
several choices of $Z'$ mass;  the width of these
distributions is mostly due to jet energy resolution and the multiple
combinations of jet-parton assignments, rather than the natural width
of the $Z'$, which is predicted to be much smaller~\cite{ourzp}.  Backgrounds, in
which no resonance is present, 
 have a broad, smoothly decreasing distribution at low $m_{t\bar{t}j}$, while a signal
 would be reconstructed near the resonance mass.

We consider several sources of systematic uncertainty on the predicted background
rates and distributions, as well as on the expectations for a
signal. Each systematic uncertainty affects the expected sensitivity
to a signal, expressed
as an expected cross-section upper limit in the no-signal
assumption. The dominant systematic uncertainty is the jet energy
scale (JES) uncertainty~\cite{jes}, followed by theoretical uncertainties on the
cross sections of the background processes.  To probe the description
of the additional jet, we compare our nominal $t\bar{t}$ model to one
generated by {\sc mc@nlo} and take the full difference as a systematic
uncertainty.  We also consider systematic uncertainties associated with the description of
initial- and final-state radiation~\cite{rad}, uncertainties in the efficiency of
reconstructing leptons and identifying $b$-quark jets, and uncertainties in the contribution
from multiple proton interactions.  In addition, we consider
a variation of the $Q^2$ scale of $W$-boson+jet events in {\sc algpen}. In each case, we treat
the unknown underlying quantity as a nuisance parameter and measure the distortion of the
 $m_{t\bar{t}j}$ spectrum for positive and negative fluctuations of the
 underlying quantity. Table~\ref{tab:syst} lists the
 contributions of each of these sources of systematic uncertainty to the
yields.

\begin{table}[!h]
\caption{Contributions to the systematic uncertainty on the expected numbers of events for the two main background processes, the total background yield,  and
  an example $500$ GeV$/c^2$ resonance signal with an assumed
  total cross section of 300 fb.}
\begin{ruledtabular}
\begin{tabular}{lcccc}
 Process & $t\bar{t}$ & $W$-boson+jets &  Total bg.& $Z'$ \\ \hline
Yield & 550 & 79& 670 & 102\\  \hline
JES & $17\%$ & $15\%$ & $16\%$ & $9\%$ \\
Cross section & $10\%$ &  $30\%$  & $12\%$ & -\\
$t\bar{t}$ generator & $6\%$ &  -  & $5\%$ & -\\ 
Gluon radiation & $6\%$ & -& $5\%$  & $4\%$ \\
($e/\mu$, $b$-jet) ID eff.  & $5\%$ & $5\%$ & $5\%$ & $5\%$ \\
Mult. interactions & $3\%$ & $2\%$ & $3\%$ & $2\%$\\
$Q^2$ scale& - & $19\%$  &$2\%$  & - \\ \hline
Total syst. uncert.& $22\%$ &  $39\%$ & $22\%$ & $11\%$\\ 
\end{tabular}
\end{ruledtabular}
\label{tab:syst}
\end{table}

\begin{figure}[htp]
\includegraphics[width=0.9\linewidth]{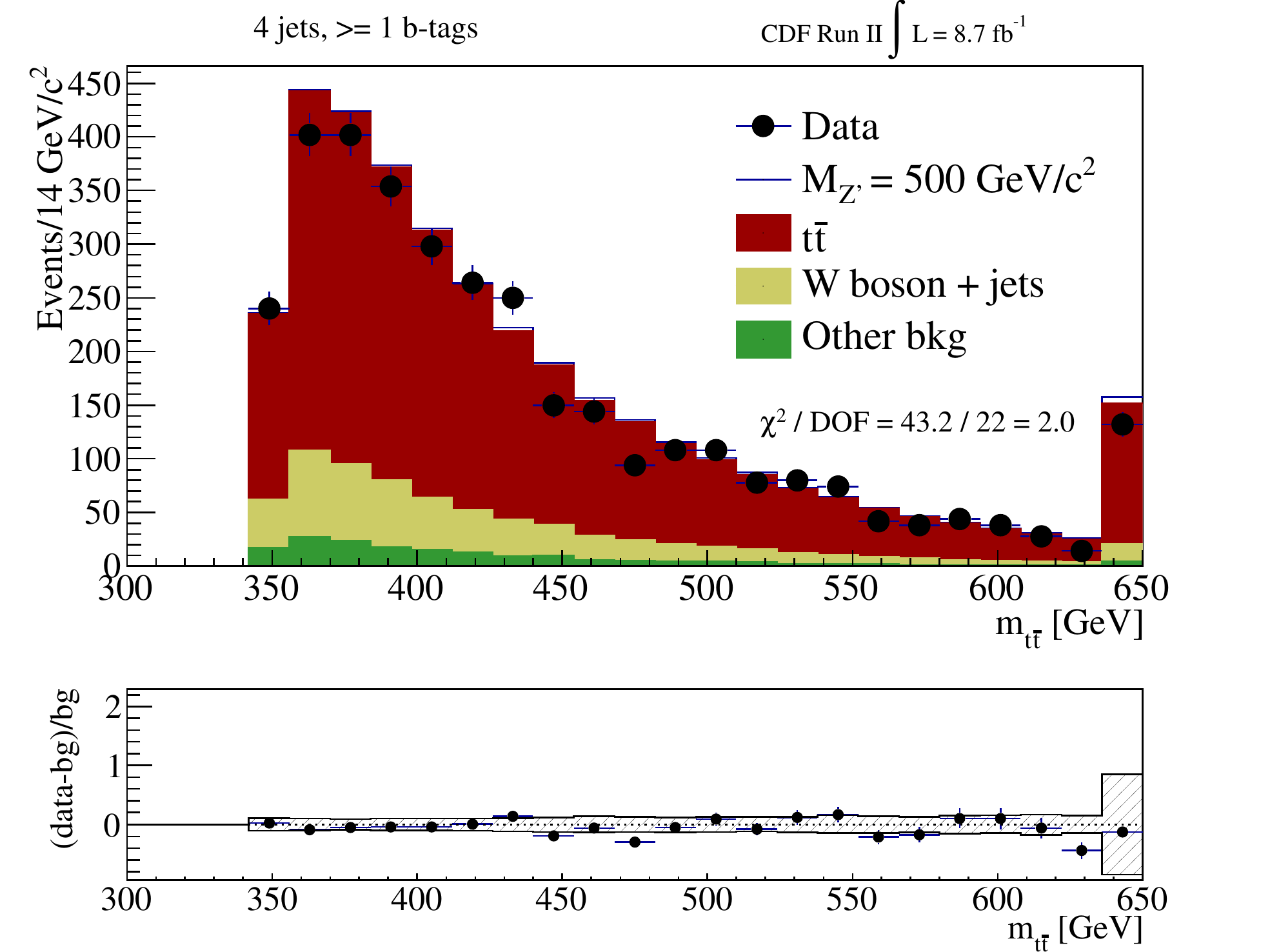}
\includegraphics[width=0.9\linewidth]{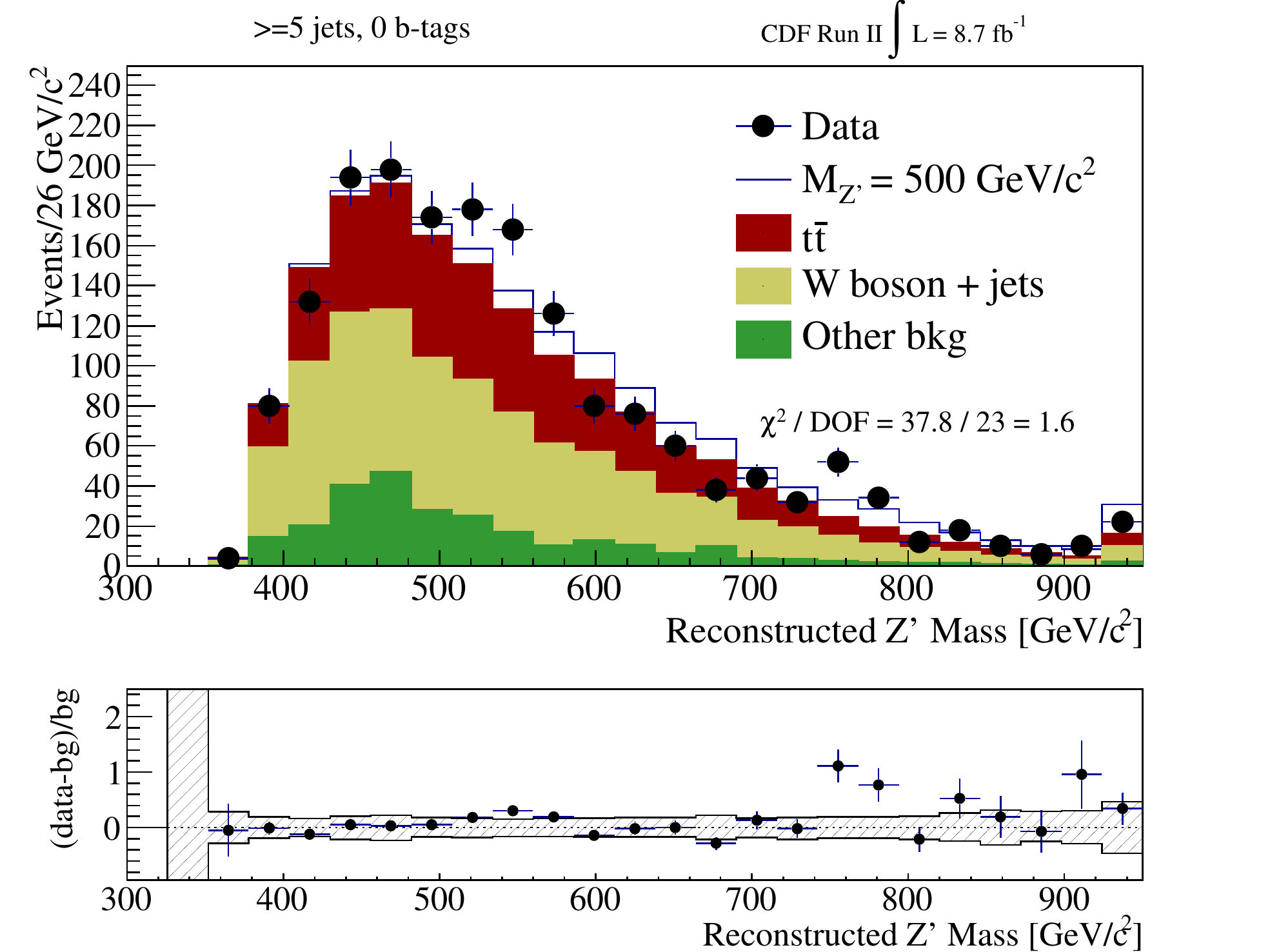}
\includegraphics[width=0.9\linewidth]{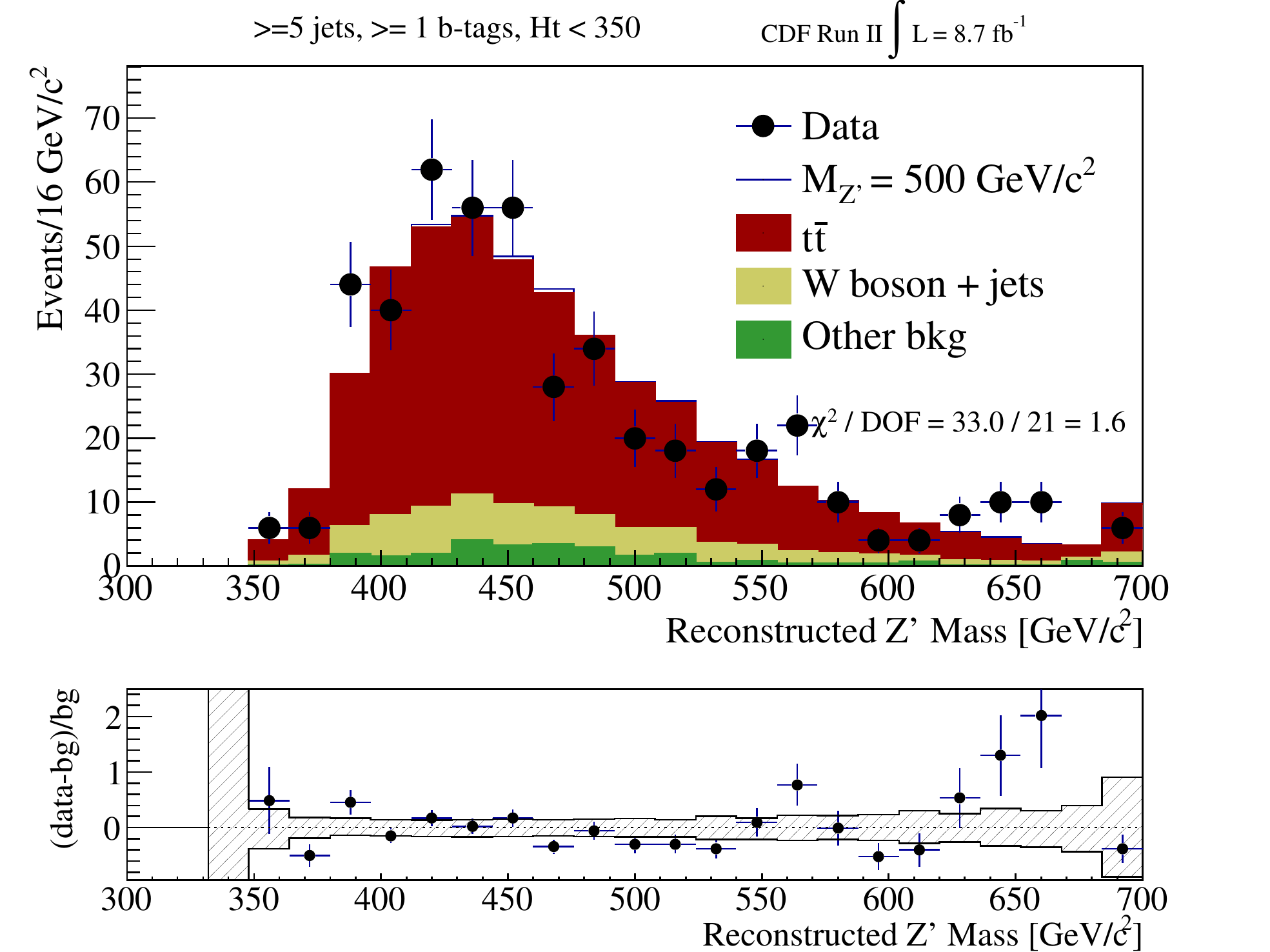}
\caption{Distribution of events versus reconstructed $t\bar{t}$ or $t\bar{t}j$  invariant mass  for observed data and expected backgrounds in three
control regions. Top, reconstructed $t\bar{t}$ invariant mass in
events with exactly four jets and at least one $b$-tag. Center,
reconstructed $t\bar{t}j$ invariant mass in events with at least five
jets and exactly zero $b$-tags. Bottom, reconstructed $t\bar{t}j$ invariant mass in events with at least five
jets at least one $b$-tag and \mbox{$H_{T}<350$} GeV. The lower panels give the relative
  difference between the observed and expected distributions; the
  hatched areas show the combined statistical and systematic
  uncertainties of the expected background. A comparison of the
  observed data and expected backround is provided by the $\chi^2$ calculation.  }
\label{fig:ctrl_region}
\end{figure}

We validate our modeling of the SM backgrounds in three
background-dominated control regions. The $t\bar{t}$ background is
validated in events with exactly four jets and at least one $b$ tag. We
validate $W$-boson+jets backgrounds in events with at least five jets and no
$b$ tags. Finally, modeling of SM $t\bar{t}$ events with an additional
jet is validated by examining a signal-depleted region with at least five
jets, at least one $b$ tag and $H_{T}$, the
scalar sum of lepton and jet transverse momenta, less than $350$ GeV$/c$. As shown in Fig.~\ref{fig:ctrl_region},  the backgrounds are well modeled within systematic uncertainties.

\begin{figure}[!h]
\includegraphics[width=0.9\linewidth]{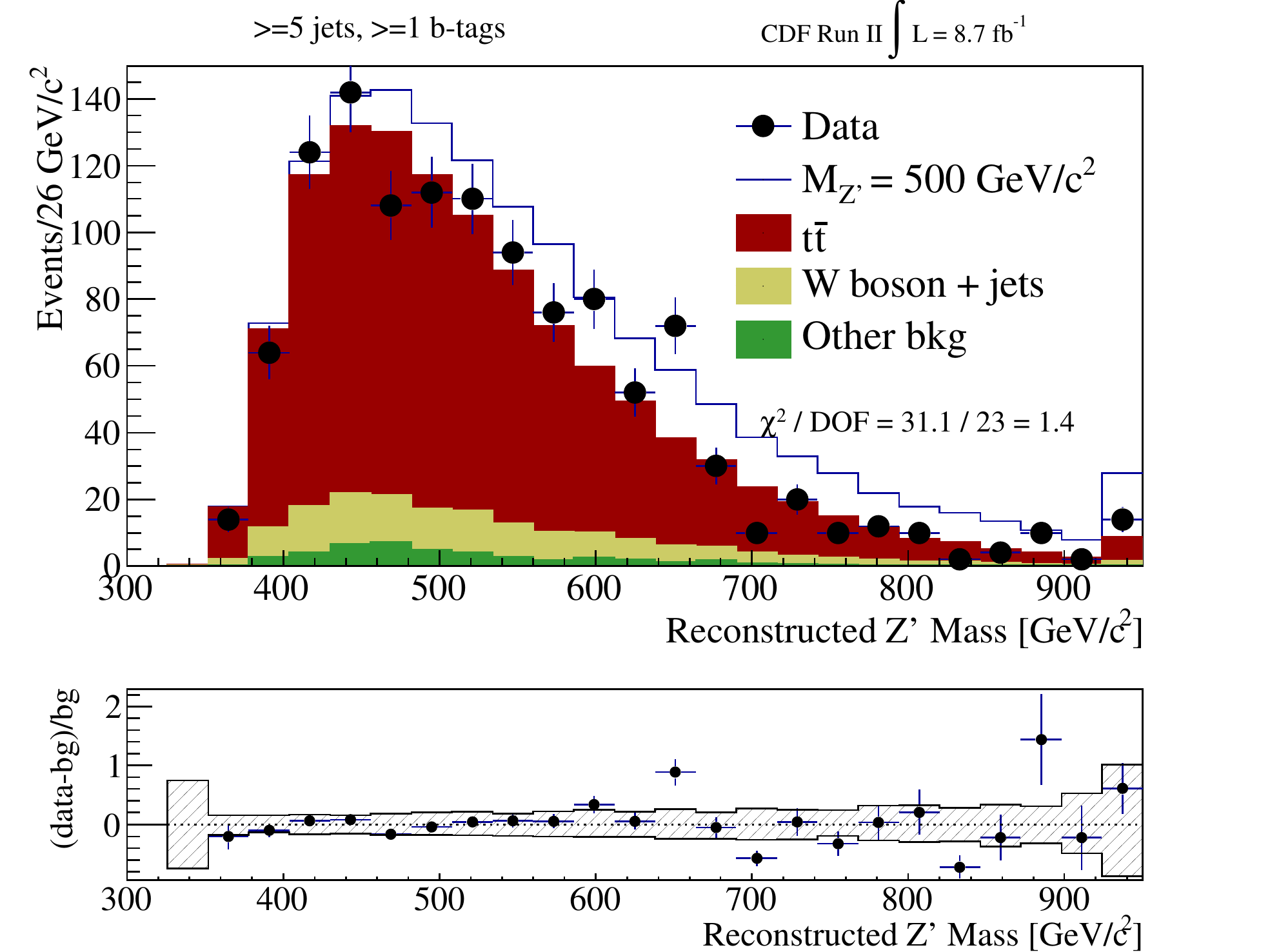}\\
\caption{Distribution of events versus reconstructed $t\bar{t}j$ invariant mass,
  $m_{t\bar{t}j}$, for observed data and expected backgrounds in the signal region.  A signal
  hypothesis is shown, assuming a total cross section of 300 fb.  The lower panel gives the relative
  difference between the observed and expected distributions; the
  hatched area shows the combined statistical and systematic
  uncertainties of the expected background. A comparison of the
  observed data and expected backround is provided by the $\chi^2$ calculation.}
\label{fig:sig_region}
\end{figure}

\begin{figure}[!h]
\includegraphics[width=0.7\linewidth]{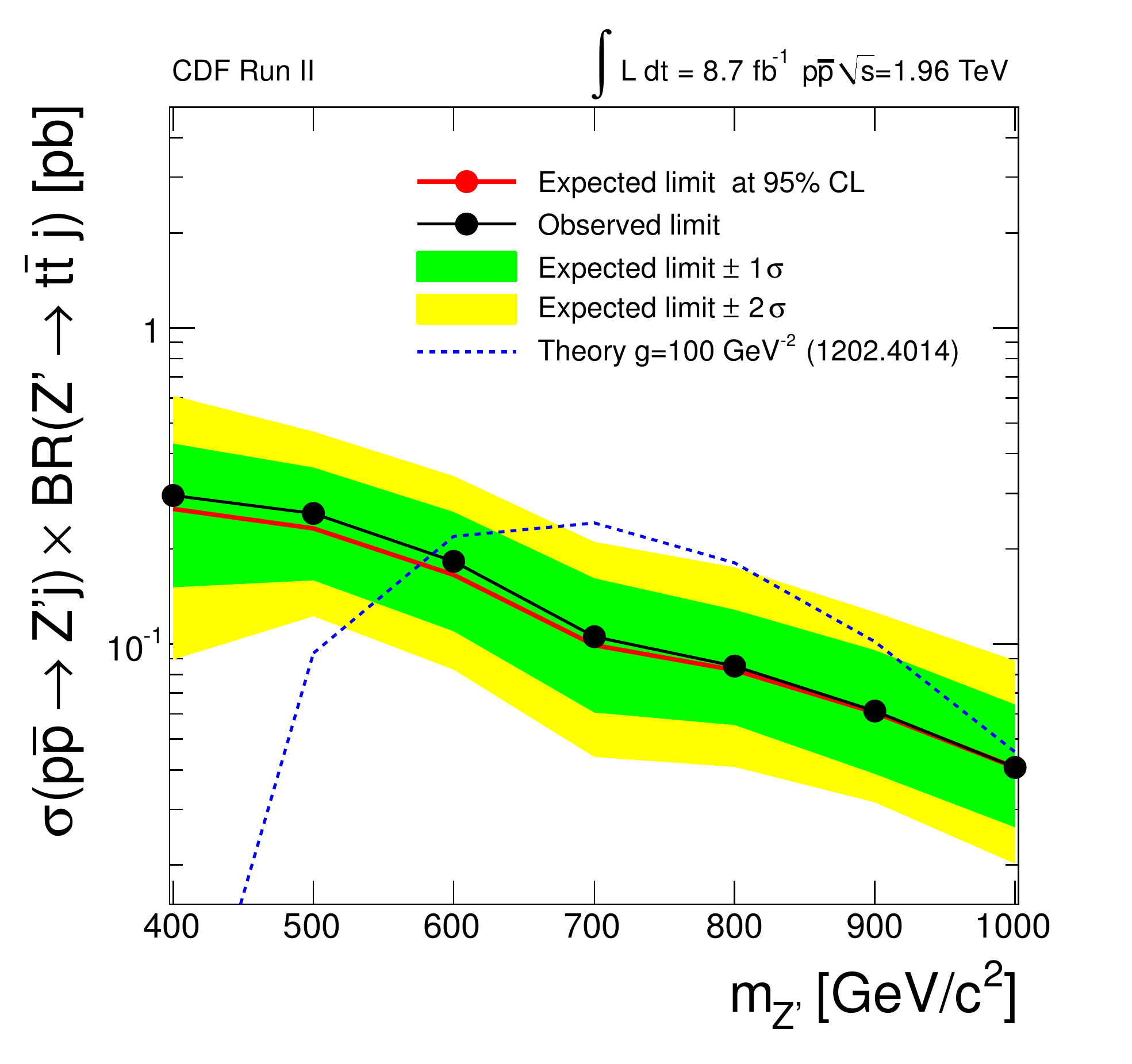}\\
\includegraphics[width=0.7\linewidth]{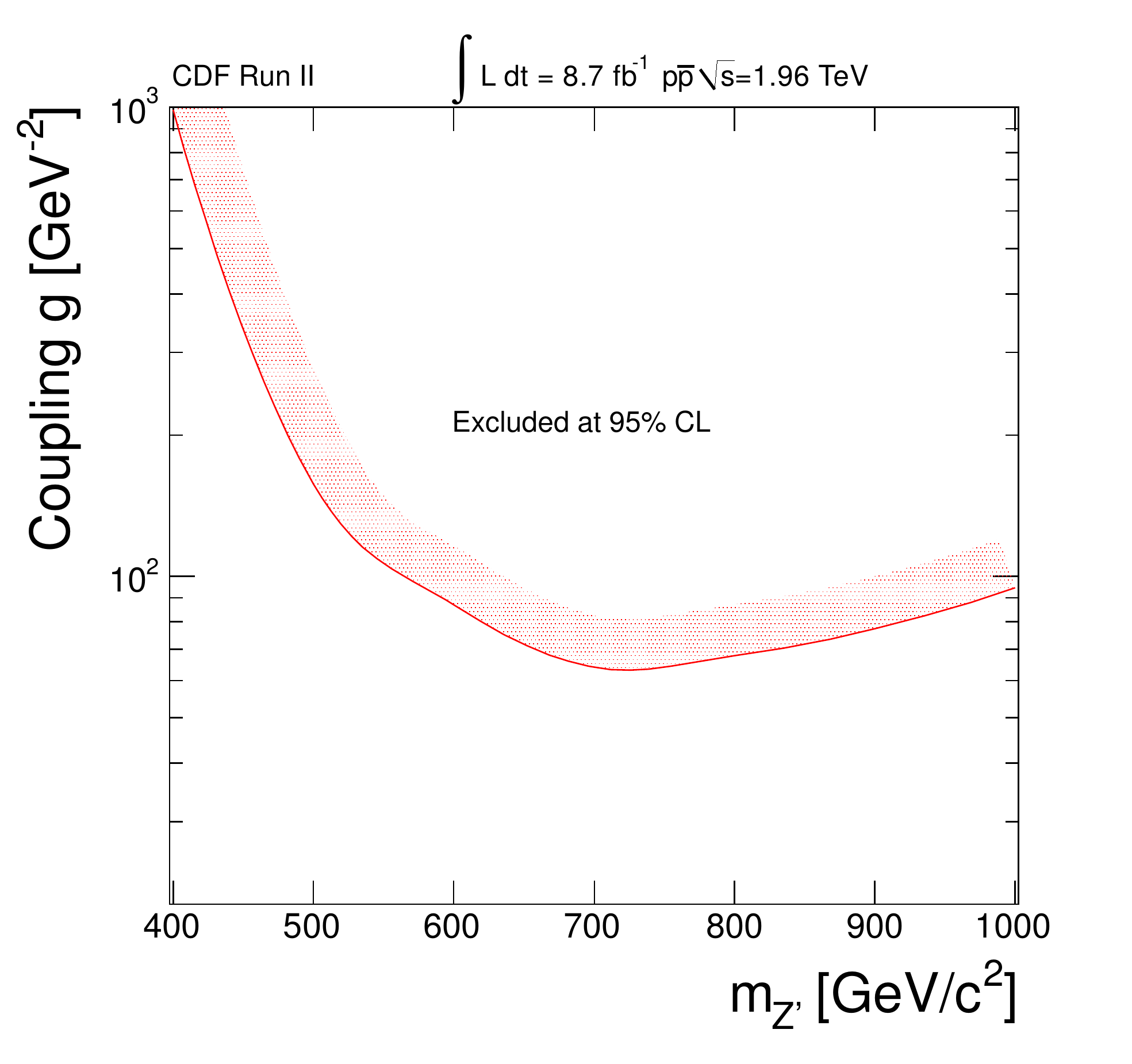}\\
\caption{Top, upper limits at 95\% C.L. on $t\bar{t}+j$ production via a heavy new
  resonance $Z'$, as a function of the resonance mass.  Bottom, limits
  on the coupling $g$ in the $Z'$ theory~\cite{ourzp} versus resonance mass.}
\label{fig:limits}
\end{figure}

Figure~\ref{fig:sig_region} shows the observed distribution of events
in the signal region 
compared to possible signals and estimated backgrounds.  At each $Z'$
mass hypothesis, we fit the
most likely value of the $Z'$
 cross section by performing a
binned maximum-likelihood fit of the $m_{t\bar{t}j}$ distribution, allowing for
systematic and statistical fluctuations via template
morphing~\cite{morph} of the signal and background distributions.  No evidence is found for the presence of top-quark-pair$+$jet resonances
in $t\bar{t}j$ events, so we set upper limits on $Z'$ boson production
 at 95\% confidence level
 using the CLs method~\cite{MCLimit}, without profiling the systematic uncertainties.
 The observed limits are
 consistent with expectation for the background-only hypothesis. The upper limits on the cross section are
 converted into limits on the coupling factor $g$ in the $Z'$
 gluon vertex~\cite{ourzp} (Fig.~\ref{fig:limits} and
 Table~\ref{tab:lim}) in order to relate the observed limits to the
 theoretical prediction. A coupling which is much larger than unity
 would make the theory non-perturbative.

\begin{table}
\caption{ For each $Z'$ mass hypothesis, the expected and observed
  limits at 95\% C.L. on the production cross section times branching
  ratio, the theoretical prediction for coupling $g=100$ GeV$^{-2}$, and the
  limit on $g$.}
\label{tab:lim}
\begin{tabular}{lrrr}
\hline\hline
$Z'$ mass\ \ \ \  & Exp (obs) limit\ \ \  & Theory \ \ \ 
 & Limit on $g$\\
(GeV$/c^2$) &  on $\sigma$ (pb) & \ \ 
 $\sigma$ (pb) & (GeV$^{-2}$)\\
\hline
400     & 0.27  (0.30)         & 0.003 & 987\\
500     & 0.23  (0.26)         &  0.09& 157\\
600     & 0.17  (0.18)         &  0.22 & 87\\
700     & 0.10  (0.11)         &  0.24 & 64\\
800     & 0.083  (0.085)         & 0.18 & 68\\
900     & 0.061  (0.061)         &  0.10 & 77\\
1000    & 0.041  (0.041)         &  0.05 & 94\\
\hline\hline
\end{tabular}
\end{table}

In conclusion, we report on the first search for top-quark-pair$+$jet resonances in
$t\bar{t}j$ events. Such resonances are predicted by various
extensions~\cite{ourzp} of the standard model and their existence is poorly
constrained experimentally. For each accepted event, we reconstruct the 
resonance mass ($m_{t\bar{t}j}$), and find the data to be consistent with SM
background predictions. We calculate 95\% C.L. upper limits  on
the cross section of such resonance production from $300$ fb to $40$ fb
 for $Z'$ masses ranging from $400$ GeV$/c^2$ to $1000$ GeV$/c^2$ and
 interpret the limits in terms of a specific
physics model.  These limits constrain a small portion of the
model parameter space. Analysis of collisions at the Large Hadron
Collider may further probe the remaining allowed regions.

We thank the Fermilab staff and the technical staffs of the participating institutions for their vital contributions. This work was supported by the U.S. Department of Energy and National Science Foundation; the Italian Istituto Nazionale di Fisica Nucleare; the Ministry of Education, Culture, Sports, Science and Technology of Japan; the Natural Sciences and Engineering Research Council of Canada; the National Science Council of the Republic of China; the Swiss National Science Foundation; the A.P. Sloan Foundation; the Bundesministerium f\"ur Bildung und Forschung, Germany; the Korean World Class University Program, the National Research Foundation of Korea; the Science and Technology Facilities Council and the Royal Society, UK; the Russian Foundation for Basic Research; the Ministerio de Ciencia e Innovaci\'{o}n, and Programa Consolider-Ingenio 2010, Spain; the Slovak R\&D Agency; the Academy of Finland; and the Australian Research Council (ARC).


\begin{thebibliography}{99}
\bibitem{coordinates}
CDF uses a cylindrical coordinate system with
  the $z$ axis along the proton beam axis. For a particle or a jet, pseudorapidity is $\eta
  \equiv -\ln(\tan (\theta/2))$, where $\theta$ is the polar angle
  relative to the proton beam direction, and $\phi$ is the azimuthal
  angle  while transverse momentum $p_T = |p|\ \sin\theta$, and the
  transverse energy $E_T = E\ \sin\theta$.
\bibitem{metdef} 
Missing transverse momentum, $\missET$, is defined as the magnitude of
the vector $ -\sum_i E_T^i \vec n_i$ where $E_T^i$ are the magnitudes
of transverse energy contained in each calorimeter tower $i$, and
$\vec n_i$ is the unit vector from the interaction vertex to the
projection of the calorimeter-tower centroid in the transverse ($x,y$) plane.
\end{thebibliography}

\end{document}